# Beyond the Expiry Date: Uncovering Hidden Value in Functional Drink Waste for a Circular Future

Yiying He [1], Zhiqiang Zuo [1,2*], Yianni Alissandratos [3], Penny Willson [3], Shameem Kazmi[3], Alex P. S. Brogan [4,*], Miao Guo [1,*]

[1] Department of Engineering, King's College London, London, WC2R 2LS, UK

[2] Water Research Centre, School of Civil and Environmental Engineering, University of New South Wales, Sydney, NSW 2052, Australia

[3] Britvic Soft Drinks Ltd., Carlsberg Britvic, Breakspear Park, Breakspear Way, Hemel Hempstead, Herts, HP2 4TZ

[4] Department of Chemistry, King's College London, Britannia House, 7 Trinity Street, London, SE1 1DB, UK

**Corresponding author**: miao.guo@kcl.ac.uk (Miao Guo), alex.brogan@kcl.ac.uk (Alex Brogan), zhiqiang.zuo@unsw.edu.au (ZhiQiang Zuo)

**Abstract**

Expired functional drinks have great valorisation potential due to the high concentration of organic molecules present. However, detailed information of the resources in these expired functional drinks is limited, hindering the rational design of a recovery system. To address this gap, we present here a study that comprehensively characterises the chemical composition of functional drinks and discus their potential use as feedstocks for biomethane production. The example functional drinks were abundant in sugars, organic acids, and amino acids, and were especially rich in glucose, fructose, and alanine. Our studies revealed that functional drinks with high COD values that corresponded to high proportions of sugar and organic acid and low proportions of sorbitol and amino acids could realise profitable recovery through anaerobic digestion, with a minimum biomethane yield of 11.72 mL $CH_4$ / mL drink. To assess utility further we also examined the dynamic composition of functional drinks up to 16 weeks (at 4 °C) after expiration to capture the shift in resources during deterioration. In doing so, we identified 4 distinct periods of carbon resource variation: 1) chemically stable period, 2) sorbitol degradation period, 3) sugar degradation period, and 4) acidification period. Based on the time-course biomethane production experiments for expired functional drinks, the optimal operating time window for biomethane production from drinks without ascorbic acid would be after sorbitol degradation period in terms of its economic performance through convenient natural deterioration. Therefore, this comprehensive study on dynamic chemical composition in expired functional drinks and their biomethane production potential could facilitate a rational design of resource recovery system for soft drink field.

# 1 Introduction

The growing emphasis on sustainability is driving a paradigm shift from a linear economy to a circular economy. In this regard, resource recovery from wastewater is considered as critically important in developing a 'zero waste' society and economy (Mirabella et al., 2014). The soft drink market is one of the largest beverage market in UK with approximately 7.9 billion litres consumed in 2023 (Ridder, 2024). Soft drink manufacturing wastewater and expired drinks constitute the two major waste streams (Marcel et al., 2023; Seng et al., 2005). In 2021-2022, the UK industry generated more than 19.75 billion litres of manufacturing wastewater ,and more than 241 thousand tonnes of expired drinks (Marcel et al., 2023; Ramirez Camperos et al., 2004). Whilst the reutilisation of soft drink manufacturing wastewater has been widely studied (Nweke et al., 2015; Peixoto et al., 2011; Ullah et al., 2015), the sustainable treatment and valorisation of expired drinks remain underexplored. According to the Waste and Resources Action Programme (WRAP), the disposal of expired drinks contributes over 650,000 tonnes of $CO_2$ equivalent greenhouse gas emissions annually (Marcel et al., 2023). Under this context, it is crucial to alleviate the environmental burden of expired drinks by developing suitable resource recovery strategies.

With the growing attention on promoting a healthy diet globally, the production of functional drinks, *i.e.* beverages formulated from fruit and vegetables to offer added health benefits, has shown an increasing trend in recent years (Ullah et al., 2015). However, unlike conventional non-carbonated beverages (e.g. non-squeezed fruit juices) or carbonated soft drinks (e.g. colas), functional drinks suffer from short shelf lives of typically 30 days at refrigerated temperatures (Bueno et al., 2021; Kokwar et al., 2022; Kumar R, 2015) compared to 6-12 months at room temperature. Approximately 95% of expired drinks are directly discharged into sewer systems (Marcel et al., 2023). As such, although expired drinks can be treated together with municipal wastewater, the potential carbon and nutrient resources contained in expired functional drinks is wasted. Therefore, recovering these resources represents a sustainable opportunity to enhance circularity with the soft drink sector and reduce its overall environmental footprint.

One of the critical challenges hindering the systematic design of resource recovery from expired functional drinks lies in their complex and dynamic chemical components. Previous studies have focused on chemical characterisation of fresh soft drinks rather than functional drinks. For instance, fresh non-carbonated fruit juices were reported to contain an overall concentration of 12 - 160 g·$L^{-1}$ for sucrose, fructose, and glucose (Ashurst, 2016) and a varying concentration of 520-4700 mg·$L^{-1}$ for free amino acids (Asadpoor et al., 2014; Zeng et al., 2015). For carbonated soft drinks, sucrose is the main component at a level of 100 g·$L^{-1}$. In comparison to these 2 typical fresh soft drinks, the investigation on chemical composition of functional drinks is largely limited (Supplementary Table S1). Moreover, expired functional drinks can undergo substantial composition changes after their expiry date due to either biological (Shankar et al., 2021) or chemical degradation (Chen et al., 2001). These processes reduce the concentration of original constituents while generating secondary metabolites such as ethanol, carbon dioxide, and organic acids (Shankar et al., 2021). Such transformation profoundly affects the resource recovery potential of these expired functional drinks. To the best of our knowledge, systematic data on the post-expiration composition of functional drinks are lacking, representing a significant knowledge gap that constrains the development and optimisation of valorisation technologies.

Therefore, to fill the knowledge gap around the resource recovery potential of expired functional drinks, this study correlates biogas production to an in-depth analysis of the chemical composition in expired functional drinks. Specifically, we first conducted a general analysis of the organic chemical composition of 5 different functional drinks and monitored biomethane production potential. Interestingly, the measured resource recovery was lower than expected based on crude COD estimates. To investigate this discrepancy, we then performed targeted chemical analyses, revealing the variation in drink chemical composition uniquely influenced on microbial community development during anaerobic digestion, thereby affecting biogas yields. Finally, to further explore the impact of storage, we monitored chemical composition of expired functional drink over16 weeks (at 4 °C) and correlated these changes with biomethane production performance to determine the time limit for profitable valorisation. Overall, this study provides the first comprehensive evaluation of expired functional drinks as a substrate for resource recovery, demonstrating that their complex and evolving chemical makeup constrains both anaerobic digestion efficiency and the window of viable utilisation post-expiry. Collectively, our findings identified a critical recovery window for this underutilised waste stream and offer an operational framework to promote more circular and sustainable soft-drink economy.

## 2 Materials and methods

### 2.1 Chemical composition analyses in functional drinks

5 different functional drinks samples were provided by Carlsberg Britvic (named as A, B, C, D, E) and characterised. For each sample, the 500 mL drink was transferred from plastic bottle to a clean 800 mL flask and continuously mixed by magnetic stirrer. For total fraction of functional drink, 15 mL homogenous drink sample was extracted to 3 clean 15 mL Falcon tubes. For soluble and solid fraction of functional drink, firstly 50 mL of homogeneous total fraction of functional drink was transferred to 3 clean 50 mL Falcon tubes using 5 mL pipette (Tacta®, Sartorius), then centrifuged at 3000 rpm for 10 minutes at 4 °C. The supernatant was transferred to clean 50 mL Falcon tubes, while the remained solids was collected in 2 mL Eppendorf tubes, namely solid fraction of functional drink. The centrifugation of supernatant was repeated until no evident solid was contained and the volume of solid-free supernatant was recorded. The final solid-free supernatant was named as the soluble fraction of functional drink. Total, soluble, and solid fraction of functional drink were used for a comprehensive characterisation as detailed in Section Analytical methods, including general chemical characteristics *e.g.* COD, detailed chemical composition which quantified individual chemical species, and microbial analyses.

## 2.2 Time-course experiment to determine functional drink stability

We designed a controlled time-course experiment with low-temperature preservation conditions to monitor chemical stability of samples and understand dynamic chemical composition change over time; in experimental design, temperature and time was treated as the controlled variable and independent variable respectively, whereas chemical composition metrics were dependent variable. Time-course experiment was conducted for 5 functional drinks from Carlsberg Britvic over 16 weeks after their expiration date. The chemical characteristic parameters were monitored at 8 time points (0, 1, 2, 3, 4, 8, 12, 16 weeks after expiry date; 0 week as benchmark). For each time points, 3 replicate sample bottles were randomly selected to represent each drink; samples were chemically characterised following analytical methods described in Section 2.4. All functional drink samples were kept in lab fridge with a constant temperature of 4 °C over the quality monitoring period. Based on the monitored chemical composition change in functional drinks under 4 °C storage condition, $Q_{10}$ rule (Bilbie and Ghizdareanu, 2021) was applied to estimate the deterioration trends for functional drinks stored at different storage temperatures (cool condition: 10 °C, room temperature condition: 20 °C) (ECA Academy, 2017a) (Supplementary Table S6).

## 2.3 Batch anaerobic digestion fed by functional drinks for biomethane production

Anaerobic digestion inoculum used for biomethane potential test (BMP) was sampled from a mesophilic (approximately 35 °C) anaerobic digestion food waste plant based in Cambridge, UK.

The BMP assay was conducted using 1 L batch bioreactor (The Nautilus Set, Anaero Tech) with an inoculum to substrate ratio of 1:1 (1 g VSS inoculum : 1 g COD substrate), inoculum VSS of 11 g·L$^{-1}$, and a headspace of 300 mL. The VSS of inoculum was obtained following the same step in Supplementary Text S1 Physico-chemical analyses. Operational temperature was maintained at mesophilic conditions (38 °C). Homogenous total fraction of 5 functional drinks (A-E from Carlsberg Britvic) expired for 0, 4, 8, 12, 16 weeks were used as substrates. And BMP test for each sample were conducted in triplicates. To record the produced volume of biomethane, gas meter (The Nautilus Set, Anaero Tech) was filled with 3 M sodium hydroxide to remove $CO_2$ and produced biomethane volume was automatically recorded by the gas meter. Finally, the biodegradability of each sample was determined by comparing the observed biomethane with the theoretical biomethane production following Eq. (1). Biomass samples were extracted from batch bioreactors at the start and end of the BMP test and analysed for the microbiome composition.

$$\textit{Theoretical biomethane volume (mL)} = \textit{Substrate COD (g)} \times 350 \ (\textit{mL CH}_4 \ / \ \textit{g COD}) \qquad \text{Eq. (1)}$$

## 2.4 Analytical methods

pH, total solids (TS), volatile solids (VS), total suspended solids (TSS), volatile suspended solids (VSS), total chemical oxygen demand (TCOD), soluble chemical oxygen demand (SCOD), total nitrogen (TN), soluble nitrogen (SN), nitrate, nitrite, ammonium, chloride, and sulphate were determined according to the Standard APHA methods for the examination of water and wastewater (APHA, 2005). Soluble protein concentration was detected using Pierce$^{TM}$ Bradford Plus protein assay method (Supplementary Text S1 Physico-chemical analyses). The concentrations of 12 chemical species (including glucose, fructose, sucrose, ethanol, mannitol, sorbitol, citric acid, oxalic acid, malic acid, ascorbic acid, lactic acid, and acetic acid) were measured using high performance liquid chromatography (HPLC). And the concentrations of free amino acids were quantified through pre-column derivatisation method using HPLC. All HPLC protocol details were given in Supplementary Table S2-4.

DNA was extracted from the solid fraction of functional drink and biomass samples from anaerobic digestion reactors using Fast DNA Spin Kit for Soil (MP Biomedicals, LLC, Solon, OH, USA) according to the steps described in the manufacturer's instructions. Primers 515FmodF (GTGYCAGCMGCCGCGGTAA) and 806RmodR (GGACTACNVGGGTWTCTAAT) targeting the V4 region of 16S rRNA gene were used for PCR. Detailed analysis procedures can be found in the Supplementary Text S3 (DNA Extraction, 16S rRNA Gene Sequencing and Data Analysis).

## 2.5 Evaluation of economic feasibility of functional drinks valorisation via anaerobic digestion

This analysis evaluated the economic viability of recovering energy from expired functional drinks through anaerobic digestion using 1000 $m^3$ up-flow anaerobic sludge blanket (UASB). The overall cost included an annualised capital expenditure and volume specific operating cost of UASB based on estimates from Durkin et al., (Durkin et al., 2022). The costs of energy consumption and sludge disposal were excluded from this assessment. Economic benefit was quantified based on revenue from biomethane sales. A breakeven threshold was identified, above which the biomethane yield from functional drinks resulted in a net positive economic return. All parameters applied and detailed calculation process were provided in Supplementary Table S6.

## 3 Results and discussion

### 3.1 General analyses of potential resources in functional drinks

The preliminary assessment of resource potential in expired functional drinks was based on a general characterisation of fresh functional drinks. Five functional drink samples composed of different fruit and vegetable blends (labelled A-E, sourced from Carlsberg Britvic) were investigated to reduce result bias (Fig. 1). Firstly, TS illustrated that the overall quantity of chemical resources in functional drinks ranged in 33-84 g·$L^{-1}$ (Fig. 1a). Although varying across formulations, this finding highlights that the chemical resource availability of functional drinks is substantial, consistently exceeding that of wastewater streams commonly studied for reutilization, *e.g.*, municipal wastewater (0.7 g·$L^{-1}$) (Rashidi et al., 2015), conventional beverage industry wastewater (3 g·$L^{-1}$) (Akbay et al., 2018), and high-TS food processing wastewater (7 g·$L^{-1}$) (de Sena et al., 2008). Notably, VS/TS ratios of 61-81% indicated that soluble organic molecules were the main source of carbon in functional drinks. With TCOD values of 45-130 g·$L^{-1}$, this meant that functional drinks can be characterized in the high organic strength wastewater category (>1 g·$L^{-1}$) (Hamza et al., 2016). The abundance of potentially biodegradable organic molecules in functional drinks indicated that anaerobic digestion (AD) could be a suitable pathway for valorising this waste stream, enabling energy recovery and contributing to circular waste management strategies.

The distribution of TCOD in functional drinks was studied to provide insight on wastewater management system design. The VSS/VS ratio of less than 16% for all the functional drinks suggested that the vast majority (> 80%) of organic material present was fully soluble (Fig. 1b). This result was further confirmed by SCOD/TCOD ratio of over 80% across all functional drinks (Fig. 1c). These findings suggest that potential bioprocess for expired functional drink recovery should prioritise liquid-phase management over organic particle removal methods (Amuda and

Amoo, 2007). Moreover, all the functional drinks were measured as acidic, with pH less than 5.1 (Fig. 1d). This indicated that acids were likely a significant fraction of the organic resources present in functional drinks. From a process design perspective, this would require an additional pH control step to avoid operational failure in downstream biological processes (Redzwan and Banks, 2007).

In addition to carbon resources, nitrogen resources for the 5 functional drinks were measured (Fig. 1e). TN was measured in the range of 730-960 mg·L$^{-1}$ for 4 functional drinks, whereas B showed the lowest TN of 190 mg·L$^{-1}$. Although varying largely across drinks, functional drinks were classified as a high nitrogen content stream (TN > 100 mg·L$^{-1}$) (Van Hulle et al., 2010) significantly higher than that of conventional beverage wastewater (TN < 60 mg·L$^{-1}$) (Cuff et al., 2018). This high TN in functional drinks could make it a potentially good substrate for AD, whereas an additional nitrogen dosage was usually recommended to augment AD for beverage wastewater (Boguniewicz-Zabłocka et al., 2017; Redzwan and Banks, 2007). Further analysis showed that the N distribution between functional drinks was different. Insoluble N occupied approximately 30% of TN in most of drinks, and even higher in E (50%). As for SN, the inorganic component - nitrate - was detected at a level of 150-190 mg-N·L$^{-1}$ in A and C, higher than that in other three drinks (70-100 mg-N·L$^{-1}$), followed by ammonium within 19-66 mg-N·L$^{-1}$ and soluble protein within 8-26 mg-N·L$^{-1}$ in all functional drinks. Although these 3 components explained most of the SN (98%) in B, a large proportion of SN (49-77%) in other 4 drinks remained unclear, likely to be derived from amino acid molecules. The nitrogen resources distributed in functional drinks are complex in terms of their solubility and organic property. In this case, anaerobic digestion that can valorise both soluble and insoluble, inorganic and organic nitrogen chemicals in functional drinks would be an ideal recovery pathway.

## 3.2 Biomethane production potential of functional drinks

Anaerobic digestion is a common recovery pathway in soft drink manufacturing wastewater treatment field to convert organic-rich wastewater into recovered electricity or energy (Nweke et al., 2015b). Our results indicated that the functional drinks were abundant in both organic carbon and nitrogen. As such, we assessed their biomethane generation potential using batch anaerobic digestion (AD) experiments, namely biomethane potential test (BMP). The cumulative biomethane production curves (Fig. 2a) showed that biomethane was immediately produced from functional drinks with no lag phase during BMP and reached plateau within a short period of 7 days, indicating that these functional drinks were all readily biodegradable (Filer et al., 2019). Furthermore, the biodegradability of functional drinks was quantitatively described by the COD conversion efficiency, expressed by dividing the cumulative biomethane volume by the theoretical cumulative biomethane volume (Fig. 2a). Specifically, for A, C, D, 72-75% of TCOD could be recovered to biomethane, while B and E showed higher biodegradability with a higher COD conversion efficiency of 80-82%. However, all of the functional drinks had biomethane production efficiency lower than that of the standard glucose substrate (95%). This demonstrated that although these functional drinks were rich in organic molecules, biomethane production was less efficient than expected based on the measured COD (*i.e.* theoretical biomethane production). This indicated that detailed chemical analysis is essential to provide an accurate prediction on utility and identify key molecular drivers and potential inhibitors affecting to AD performance.

To further evaluate the economic feasibility of resource recovery from functional drinks through anaerobic digestion, a profitability estimation considering the implementation of the most widely applied high-rate anaerobic digester - up-flow anaerobic sludge blanket (UASB) (Mainardis et al., 2020) - was undertaken based on the laboratory biomethane yield data (Fig. 2b). This analysis showed that when the biomethane yield from functional drinks is above 11.72 mL $CH_4$ mL drink$^-$

[1], a profitable biomethane production system could be realised. Based on this benchmark, most of the functional drinks except for D are economically feasible to be reutilised through anaerobic digestion. As for D, its low biomethane yield is due to its limited quantity of organic material that could be digested (indicated by its lowest TCOD among 5 drinks). Therefore, to enable the profitability of biomethane production from low-COD drinks, a simple method could be to mix it with other high-COD functional drinks.

To understand the underlying microbiome dynamics in AD reactors fed by expired functional drink, biomass samples were collected at the start and end of anaerobic digestion process and analysed using 16S rRNA sequencing (Fig. 2c). This analysis revealed the variations of the top 10 genus in microbial communities. Notably, *Fermentimonas*, a genus recognised as a core functional group in anaerobic digestion (An et al., 2024), exhibited a substantial increase in relative abundance from approximately 0.5% to 11%, emerging as the dominant genus by the end of the process across all functional drinks treatments. Additionally, *Candidatus_Cloacimonas* and *Sedimentibacter* became the second and third most abundant genera, with an averaged relative abundance of 10% and 7%, respectively. These taxa have been reported functional in amino acid fermentation and utilisation (Imachi et al., 2016; Zhu et al., 2021). The rapid growth of *Fermentimonas*, *Candidatus_Cloacimonas*, and *Sedimentibacter* within 7 days indicates efficient microbial utilisation of organic carbon and nitrogen from functional drinks. However, for effective biomethane production, the presence of methanogenic archaea is critical. *Methanosarcina*, a key methanogenic genus, remained at low relative abundance in anaerobic digestion, ranking 7[th] in microbial communities, with only modest increase from 1% to 5% for Drink B and to 3% in the remaining drinks. This microbial composition provided insight into the relatively higher biomethane production efficiency (82%) in B compared to the other functional drinks (< 80%) while also highlighting the lower conversion efficiency of functional drinks compared to standard glucose (Fig. 2a). Overall, whilst the organic constituents of functional drinks were readily utilised by fermentative and amino-acid-degrading microorganisms, they do not support methanogenesis as efficiently, limiting their full conversion into biomethane.

**3.3 Effect of chemical molecules in functional drinks on biomethane production by AD**

As revealed by our biomethane production from functional drinks experiments, a crude measurement of COD, which is commonly applied in the wastewater treatment field, is not sufficient to understand the resource recovery potential. Therefore, we decided to investigate this further through a detailed characterisation of functional drink composition using high-performance liquid chromatography (HPLC) to identify and quantify the specific molecules, drivers, and potential inhibitors in functional drinks affecting biomethane performance.

Sugars and organic acids are prevalent chemicals in fruits and vegetables and are common carbon sources for AD. As such, sugar and organic acid concentrations were measured in the functional drinks to provide insight into their contributions to biomethane production (Fig 3a). Glucose, fructose, and sucrose concentrations varied largely between drinks in the range of 4-15 g·L$^{-1}$, 3-30 g·L$^{-1}$, 0-35 g·L$^{-1}$, respectively. These sugars constituted a major organic carbon resource in functional drinks, occupying dominant TCOD proportion of 42-58% (Fig. 4). As for organic acids, citric acid and malic acid were found as top 2 abundant in 5 functional drinks at a level of 1-5 g·L$^{-1}$ and 2-3 g·L$^{-1}$, respectively, while oxalic acid and ascorbic acid were at trace concentrations (< 0.6 g·L$^{-1}$). Citric acid could directly serve as substrate to supply energy and carbon source for fermentative organisms (Im et al., 2021). Moreover, the ability of citric acid to stimulate microbial activity and alleviate the inhibitory effect of humic acid *in-situ* produced in anaerobic digesters could improve biogas yields and process stability (Li et al., 2023; Tang et al.,

2023). As for malic acid, its effects on AD were rarely discussed in the literature. However, some studies on alcoholic fermentation revealed that malic acid could stimulate sugar fermentation by serving as an energy source through malolactic fermentation by lactic acid bacteria under anaerobic condition (Firme et al., 1994). The generated lactic acid from malolactic fermentation is an intermediate in AD process and could then be utilised by acetogenic or methanogenic bacteria (Pipyn and Verstraete, 1981). It was discovered that drinks B and E were rich in sugars and organic acids (TCOD: 62%), corresponding well to the biomethane conversion efficiency reached 80-82%. Conversely, drinks A, C, and D had lower concentrations of sugar and organic acids (TCOD: 48-56%), meaning efficiency was reduced to 72-75% (Fig. 2a). This result suggested that sugars and organic acids in functional drinks were likely the key molecular contributors to biomethane production efficiency.

Nevertheless, although rich in easily biodegradable chemicals, the overall biomethane production (< 82%) from functional drinks was less efficient than that from standard glucose substrate (95%) (Fig. 2a). This suggested that less biodegradable chemicals, or even inhibitors, could potentially exist in the functional drinks, thus requiring further analysis. As normal components in plants which was used as raw material for functional drinks (Grembecka, 2015; Legaz and Vicente, 2004), sugar alcohol was explored in functional drinks (Fig. 3a). Sorbitol was identified in 4 functional drinks except for B, with concentrations of 2-13 g·L$^{-1}$ in A, C, D and negligible amount (< 0.7 g·L$^{-1}$) in E. Evidenced by Fynn and Rankin's work, sorbitol was found to be metabolically inert in AD (Fynn and Rankin, 1987). Therefore, the presence of sorbitol may help explain the reduced biomethane production efficiency (72-75%) observed in drinks A, C, and D.

In addition to sorbitol, drinks A, C, and D also contained high levels of amino acids (5-16% of TCOD) (Fig. 3b and 4), potentially contributing further to the low efficiency of 72-75%. Similar results were reported by (Wang et al., 2021) that amino acids provided a lower methane conversion efficiency of 2-88% lower than that of standard sugars (95%). This finding indicates that higher amino acid content could negatively affect the biomethane production efficiency of functional drinks. One potential explanation for this is because unlike glucose, which is directly and efficiently metabolised via glycolysis (Mosey, 1983), amino acid degradation involves more complex and energetically demanding catabolic pathways, *e.g.,* Stickland reaction and single amino acid fermentation (Ramsay and Pullammanappallil, 2001). These pathways often yield intermediate products *e.g.,* ammonia, short-chain volatile acids, and branched-chain compounds that may accumulate to inhibitory concentrations or not be readily converted into methane.

## 3.4 Chemical variation in expired functional drinks: carbon and nitrogen resource shift along time

Having shown that the chemical complexity of functional drinks could greatly impact the resource recovery, we also wanted to investigate whether there was an optimal window post expiry to utilise functional drinks. To investigate, we performed a time-dependent chemical characterisation to track the functional drink quality over 16 weeks under a controlled storage temperature of 4 °C (consistent to European regulation on refrigeration temperature (ECA Academy, 2017b)) to mitigate the deterioration after their shelf lives (Akpe et al., 2022; Jemimah et al., 2013).

For organic carbon resources, different expired functional drinks showed different composition variations after expiration (Fig. 5a). For drink with ascorbic acid, notably, B showed relatively stable organic carbon chemicals till week 16, which might be caused by the antimicrobial effect of ascorbic acid in B (Pandit et al., 2017; Tajkarimi and Ibrahim, 2011). The other 4 functional drinks displayed four-distinct stages of chemical stability: the chemically stable stage, the sorbitol degradation stage, the sugar degradation stage, and the acidification stage. Specifically, in week 0-4 after expiry date, organic carbon chemicals remained relatively stable (chemically stable stage), after which the concentration of organic chemicals started to diminish. In weeks 4-8, sorbitol degradation was the most significant, dropping for 62% in A and 100% in C-E (sorbitol degradation stage). In week 8-12, sugar degradation was the major activity, with fructose decreased by 16-20%, glucose by 2-5%, and sucrose by 19% (sugar degradation stage). And at week 12, deterioration products (*i.e.,* ethanol, lactic acid, acetic acid, mannitol) occurred, suggesting the starting of an acidification stage, where the organic carbon starts to shift from being predominantly sugars to being predominantly deterioration products, such as lactic acid, acetic acid, and mannitol. Till week 16, deterioration products occupied 74-88% of all organic carbon chemicals in A, C, D and 49% in E. Lactic acid and mannitol represented predominant chemicals, reaching 14-20 g·L$^{-1}$ and 8-17 g·L$^{-1}$ respectively in 4 drinks (carbon source shift stage). The organic nitrogen content variation in functional drinks was also studied (Fig. 5b). However, in this case, organic nitrogen resource variation was not as complex as that of organic carbon chemicals. Amino acid molecules were relatively stable in week 0-12, with a sharp decrease of Ala (< 96%) in week 12-16. This amino acid decrease happened at the period when sugar was largely converted into deterioration products, indicating potential fermentation microorganism growth applying amino acid as nitrogen source (Fairbairn et al., 2017).

Microbial profiling aimed to determine whether the observed degradation of organic compounds in expired functional drinks was driven by the microbial biochemical activities (Fig. 5c). B maintained a relatively stable microbial composition over time, which might be linked to the presence of ascorbic acid. The antimicrobial property of ascorbic acid could help to supress fast-growing strains (*e.g.,* lactic acid bacteria) by inhibiting their biofilm formulation (Przekwas et al., 2020). However, in the other 4 functional drinks, *Lactobacillus* was enriched from a relative abundance of 5% to more than 70% after 16 weeks post expiration, explaining the production of lactic acid. For the formation of mannitol in expired functional drinks, some studies have proposed that some lactic acid bacteria (e.g., *Lactococcus lactis, Leuconostoc pseudomesenteroides,* and *Leuconostoc mesenteroides*) is able to produce mannitol at the presence of fructose and glucose (Grobben et al., 2001). However, the growing microorganisms during drink spoilage also introduces additional challenges due to their competition with functional microbial community for waste valorisation. Thereby, further investigation would be needed to understand the effect of the microbial community dynamics in expired functional drink on fermentation biotechnologies, such as biomethane production by AD.

HPLC and microbial analyses revealed that the chemical compositions and microorganism environment in functional drinks changed dramatically in the weeks 16 after expiration. We therefore performed biomethane production experiments at each time point to assess the practicability to valorise expired functional drinks via AD and determine the optimal recovery time window (Fig. 6a). These experiments showed the variations of conversion efficiency from TCOD to biomethane in 5 functional drinks over expiration (0-16 weeks under refrigeration). For drink containing antimicrobial ascorbic acid, COD conversion efficiency of B remained stable at 81.4 ± 2.6 % in 0-16 expiry weeks, which was agreed with the stable chemical compositions in B during this period. It indicates that for functional drinks with antimicrobial chemicals, a 16-weeks operational window (stored under refrigeration) for drink collection and delivery after expiry date could still achieve maximum biomethane recovery. As for drink (A, C, D) with innegligible sorbitol (> 1.5 g $L^{-1}$ and > 3.7% of TCOD), the COD conversion efficiency showed significant increase of 7.0-10.9% after 8 weeks' natural deterioration under refrigeration, achieving the maximum level when sorbitol was mostly digested. Agreed with the report that sorbitol was metabolically inert (Fynn and Rankin, 1987), our findings further indicate that when less biodegradable sorbitol occupied certain proportion in substrates (> 3.7% of TCOD), it could largely sacrifice the overall resource recovery efficiency via AD. Interestingly, by week 16, when organic carbon source was shifted from predominantly sugars to predominantly deterioration products, *i.e.,* acetic acid, lactic acid, and mannitol, the overall COD conversion efficiency was not significantly impacted among most of expired functional drinks. Specifically, when deterioration products occupied 49-88% in A, D, E, a small efficiency variation of -3.2% to 2.24% was observed. Consequently, to maximise the resource recovery from expired functional drink without antimicrobial chemicals, the optimal time window would be after complete sorbitol degradation through convenient natural deterioration.

Notably, by comparing to the profitable yield (11.72 mL $CH_4$ / mL drink) in Fig. 6b, it was discovered that most of the functional drinks could realise profitability even after 16 weeks' expiration under refrigeration. And although their biomethane yields showing a variation of -14.32% to +46.81% along expiry time, the profitability potential for resource recovery from functional drinks via AD was still largely dependent on their TCOD value. Therefore, it is important to recognise that while the selection of an optimal time window may improve the overall profitability, it is unlikely to exert a substantial influence on the overall economic feasibility of resource recovery from functional drinks.

### 3.5 The effect of storage conditions on resource recovery from functional drinks

The storage conditions affect resource recovery from expired functional drinks by tuning the speed of chemical change in expired functional drinks, which could influence the final time point for reutilisation and finally differ the economic cost for drink management. The turning points of chemical change are highly related to the storage temperature for expired functional drinks. To provide potential insight, we (Supplementary Table S5) estimated the equivalent storage duration for expired functional drinks under typical European refrigeration, cool, and room temperature conditions (ECA Academy, 2017a) (Supplementary Table S5). The estimation showed that room temperature storage requires a 19-day degradation length for sorbitol natural digestion, while cool storage requires a longer waiting window for 37 days post-expiry as compared to the 8 weeks measured under refrigeration. Storage under room temperature and cool conditions could shorten the operational window avoiding higher economic cost associated with energy consumption and warehouse space, however raise requirements for quick operation for drink collection, delivery, and fermentation. It should also be noted that there would be a time threshold after 16 week post expiration, before when expired functional drinks could still be recovered via AD from the perspective of practicability or profitability. Therefore, it is crucial to integrate storage and drink management options into decision-making process when designing economically viable and scalable resource recovery system for expired functional drink.

## 4  Conclusion

This study provides the first systematic evaluation of expired functional drinks as a feedstock for biotechnological valorisation. Through integrated chemical profiling and biomethane potential assays, we revealed that these drinks are rich in biodegradable organic carbon and nitrogen, yet their complex and evolving composition influences energy recovery performance. The identification of four distinct post-expiry stages of carbon-source transformation offers new insight into substrate dynamics and defines an optimal operational window, after sorbitol degradation for maximising biomethane yield and economic return.

Our findings reveal that while COD quantifies total oxidisable carbon, it obscures the underlying biochemical heterogeneity that governs biodegradability and conversion efficiency. The contrasting behaviour of easily degradable sugars and less biodegradable compounds such as sorbitol and amino acids demonstrates that molecular-level characterisation, rather than aggregate chemical indices, is essential for accurately predicting recoverability and for rationally designing circular waste-to-energy systems in the beverage sector. By establishing correlates between chemical composition, storage stability, and anaerobic digestion performance, this work not only offers a mechanistic understanding of how substrate chemistry evolves during drink deterioration but also provides a practical framework for integrating beverage waste valorisation into existing AD infrastructures. While this work was conducted under controlled laboratory conditions, future research should focus on pilot-scale validation across diverse formulations, assessment of real-world variability, and integration with complementary valorisation routes such as microbial protein, biosurfactant, or biosolvent production. Incorporating AI-driven modelling to predict degradation kinetics and optimise recovery timing could further enhance process resilience and scalability.

Overall, this study advances both scientific and operational understanding of expired functional drinks as an underutilised yet high-value bioresource, providing a blueprint for the development of data-driven, circular, and economically viable biotechnologies within the beverage sector.

## 5 Figures

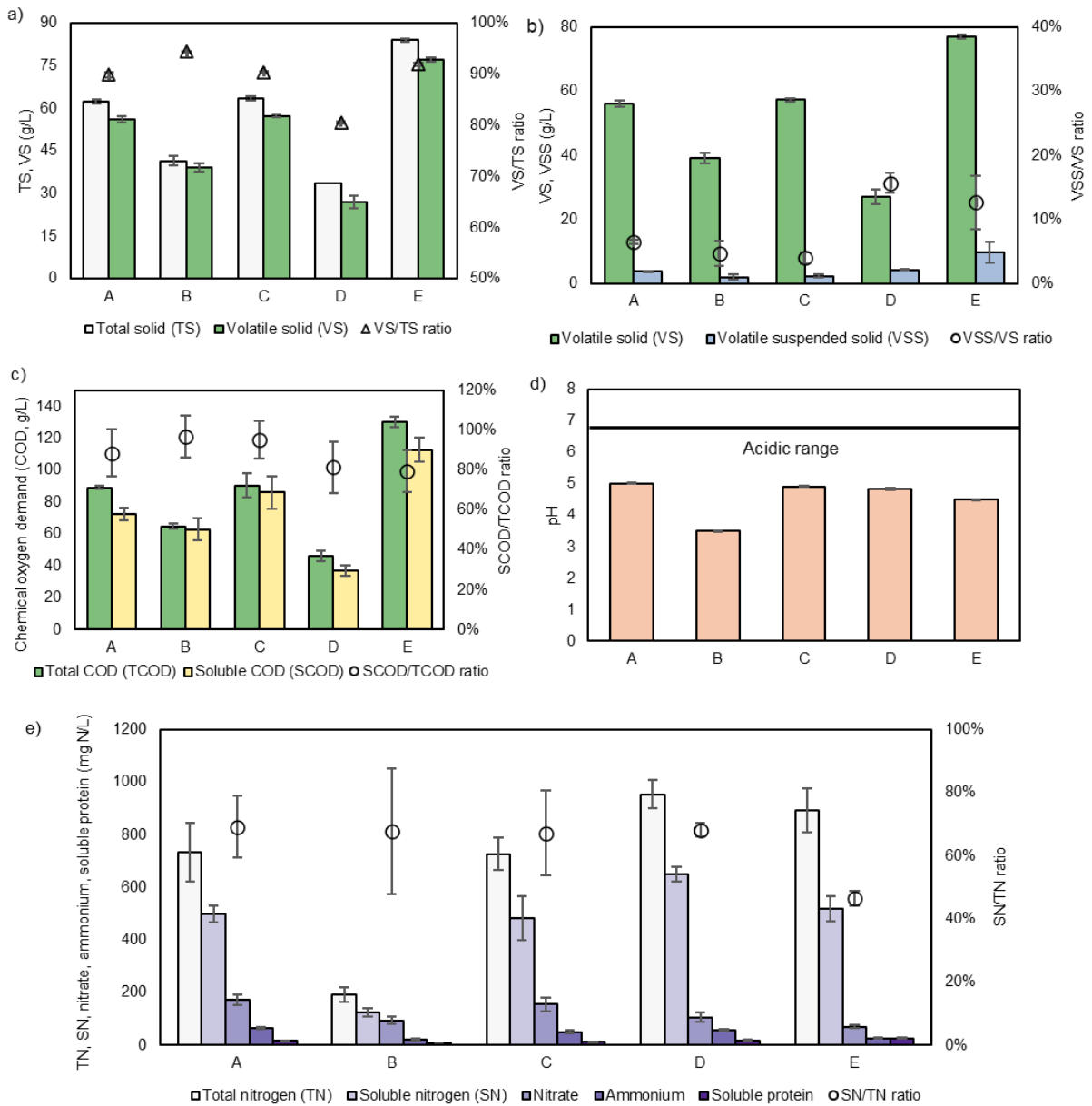

**Fig. 1.** General analysis of 5 functional drinks (A-E from Carlsberg Britvic). a) Total resource and organic resource, b) insolubility of organic resource, c) solubility of organic resource, d) acidity, and e) nitrogen distribution

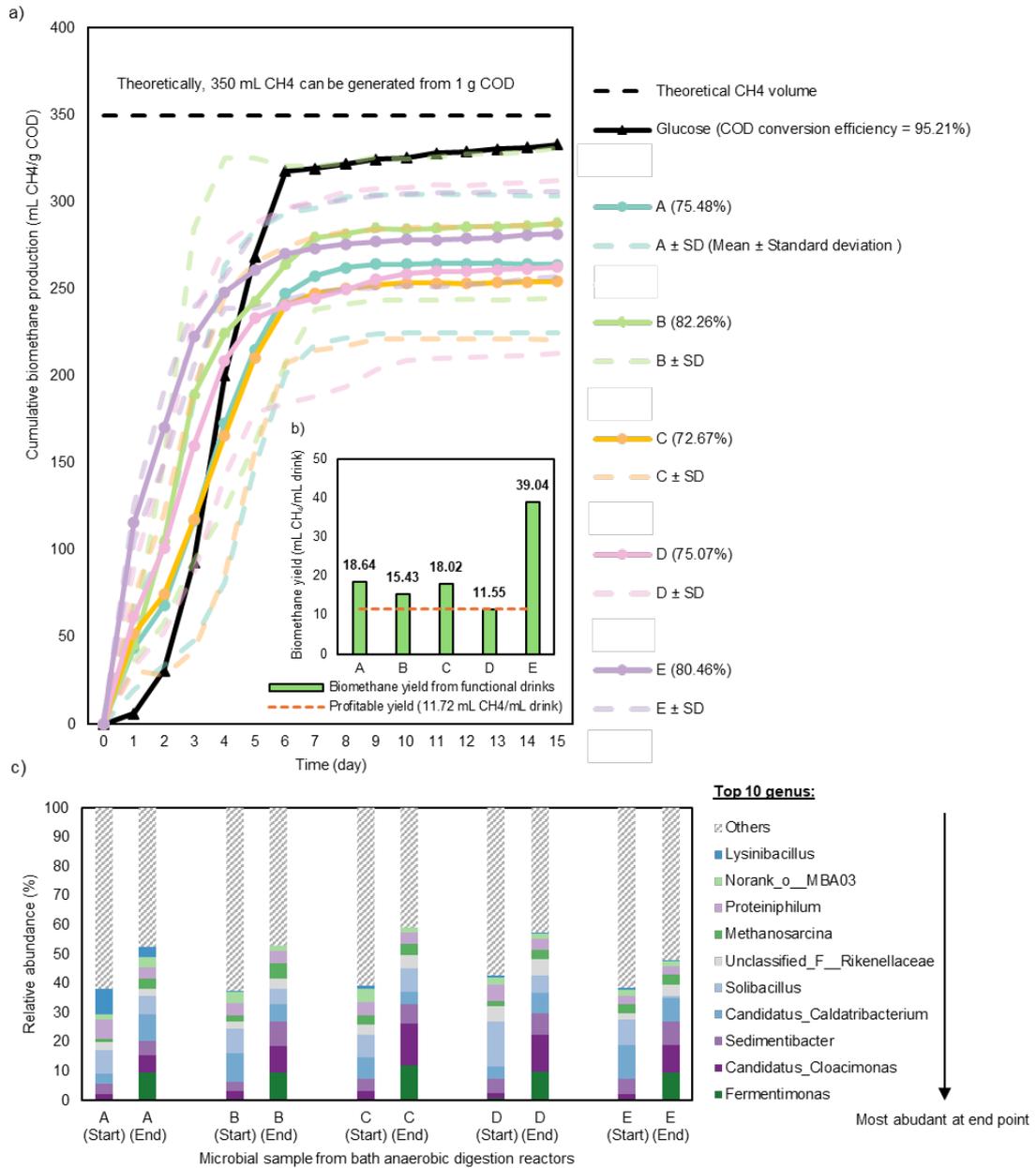

**Fig. 2.** (a) Cumulative biomethane production from functional drinks (A-E from Carlsberg Britvic), their (b) biomethane yields compared to profitable yield, and (c) variations on top 10 abundant microbial genus in biomass during their biomethane production. Biomass samples were collected at the start and end time points of batch anaerobic digestion process fed by functional drinks and analysed using 16S rRNA sequencing to capture the genus variations. (See Supplementary Fig. S1 for individual biomethane production figures for each function drink and Table S6 for profitable yield calculation).

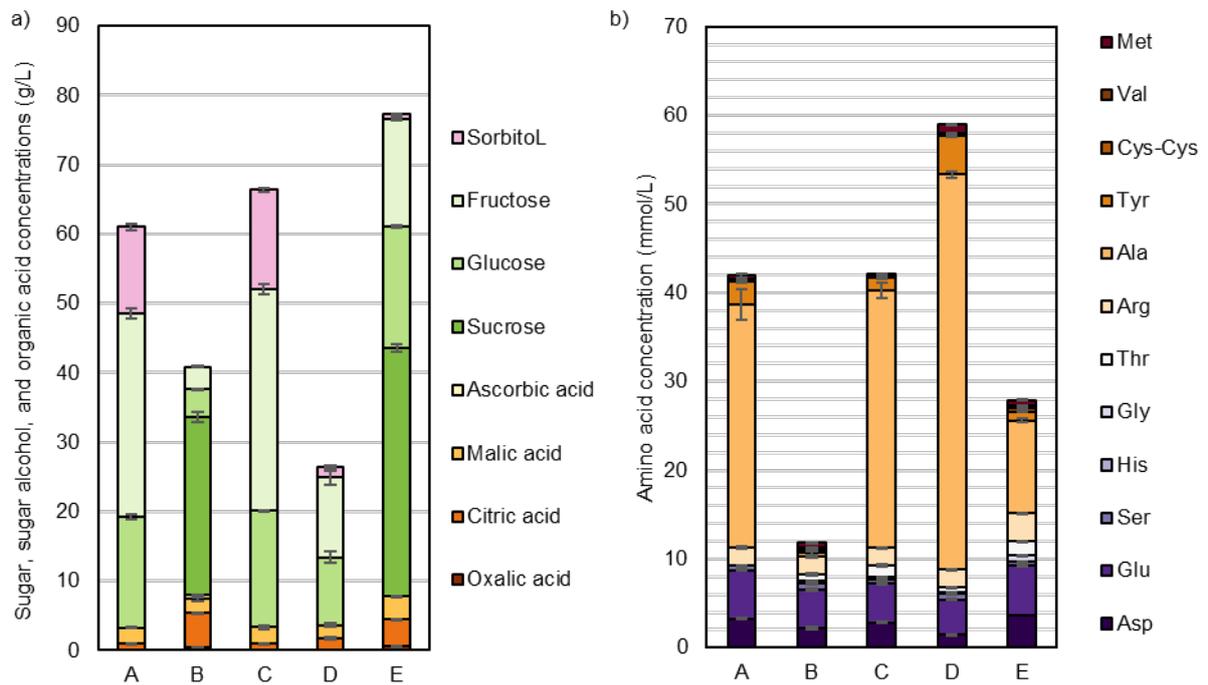

**Fig. 3.** Detailed chemical composition of functional drinks (A- E from Carlsberg Britvic). a) Organic carbon content: sugars, sugar alcohols, and organic acids and b) Organic nitrogen content: amino acids profile.

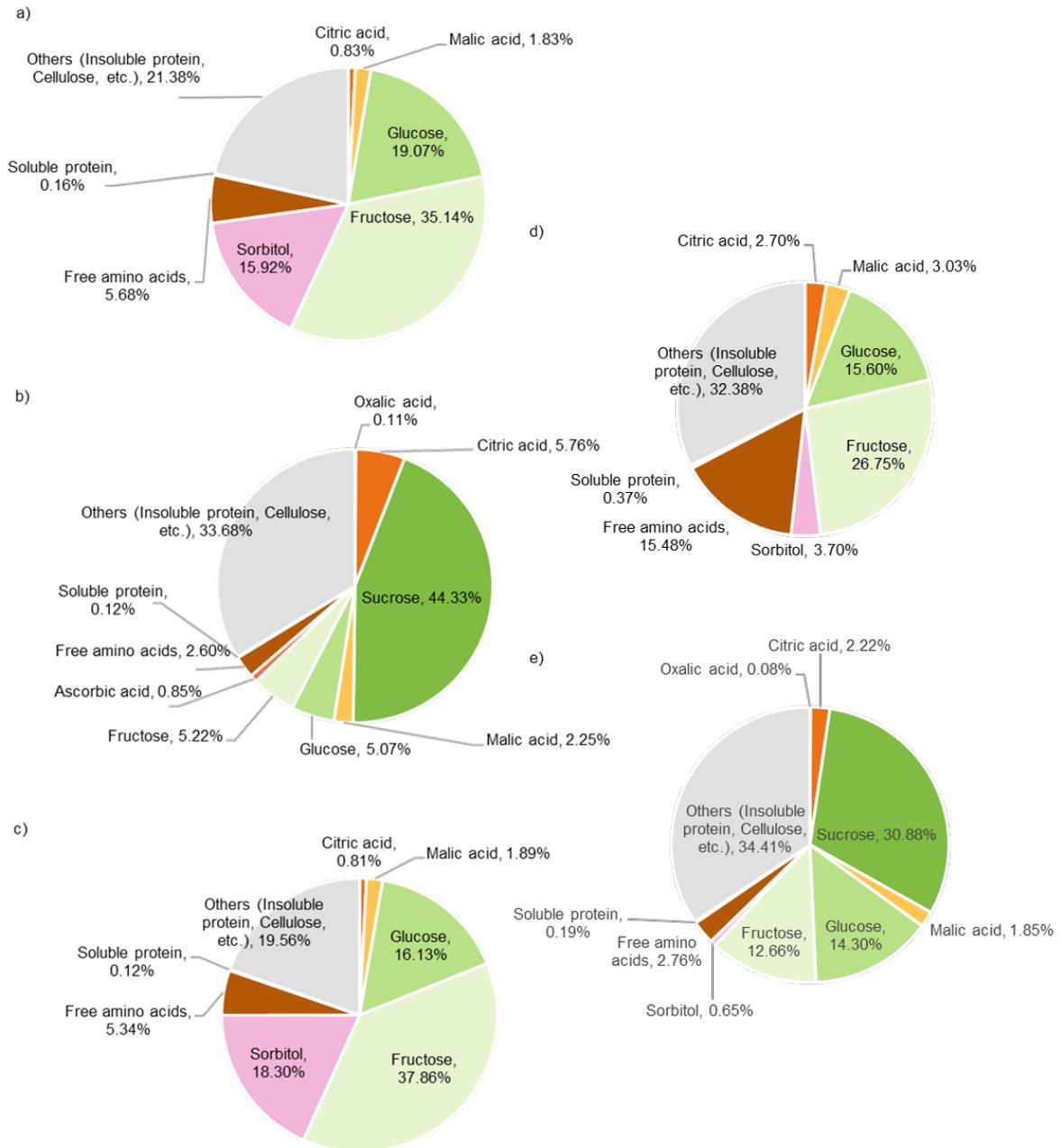

**Fig. 4.** Total chemical oxygen distribution in functional drinks (A-E from Carlsberg Britvic). a) A, b) B, c) C, d) D, e) E.

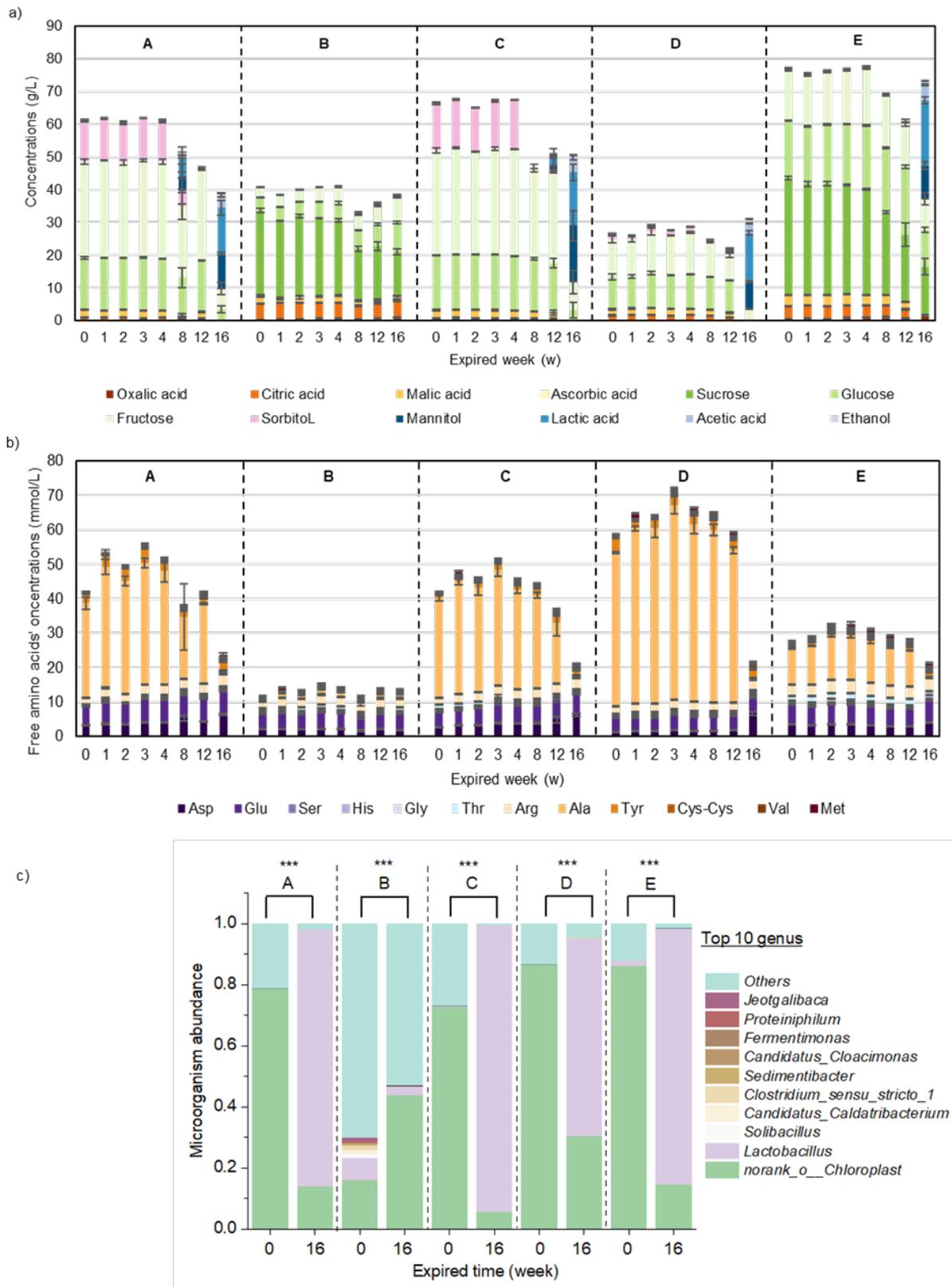

**Fig. 5.** Variation of organic carbon content (a), nitrogen content (b), and relative abundance of microbial species (c) in functional drinks (A-E from Carlsberg Britvic) expired for 0-16 weeks (Drinks kept in 4 °C fridge).

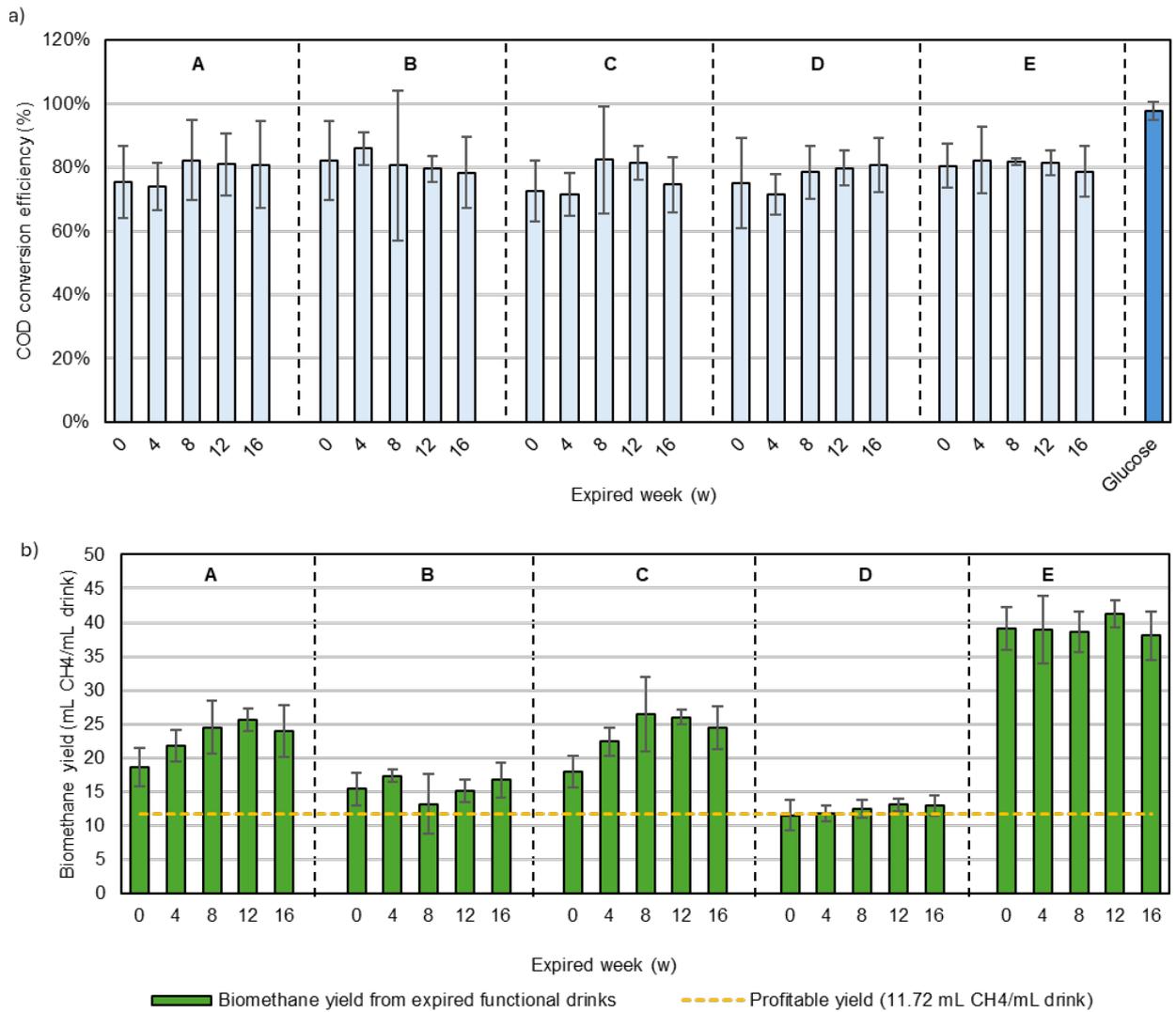

**Fig. 6.** COD conversion efficiency (a) and biomethane yields (b) of expired functional drinks expired for 0-16 weeks (A-E from Carlsberg Britvic) compared to profitable biomethane yield.

**Author Contributions**

**Declaration of Competing Interests**

The authors declare that they have no known competing financial interests or personal relationships that could have appeared to influence the work reported in this paper.

**Acknowledgements**

This study was undertaken as part of Britvic Soft Drinks Ltd.'s environmental sustainability initiatives under its ESG strategy, with a specific focus on Water Stewardship and the valorisation of post-market beverage streams. Following the acquisition by the Carlsberg Group, this work now aligns with the Carlsberg SAIL strategy ("Together Towards ZERO and Beyond"), contributing to targets related to water circularity, carbon footprint reduction, and resource-efficient innovation. The research reflects an applied commitment to circular economy principles within the functional beverage sector.

Y.H. also acknowledged the support from the Chinese Scholarship Council (CSC) and King's College London.

**Appendix. Supplementary Materials**

Supplementary material associated with this article can be found, in the online version, at xx.

**Data Availability**

Data will be made available on request.